# High-pressure phase transitions and compressibility of wolframite-type tungstates

J. Ruiz-Fuertes[1], S. López-Moreno[2], D. Errandonea[3], J. Pellicer-Porres[1], R. Lacomba-Perales[1], A. Segura[1], P. Rodríguez-Hernández[4], A. Muñoz[4], A.H. Romero[2], and J. González[5]

[1]MALTA Consolider Team, Departamento de Física Aplicada - ICMUV, Universitat de València, Edificio de Investigación, c/Dr. Moliner 50, 46100 Burjassot, Valencia, Spain

[2]CINVESTAV-Querétaro, Libramiento Norponiente No 2000, Real de Juriquilla 76230 Querétaro, México

[3]MALTA Consolider Team, Departamento de Física Aplicada - ICMUV, Fundación General de la Universitat de València, Edificio de Investigación, c/Dr. Moliner 50, 46100 Burjassot, Valencia, Spain

[4]MALTA Consolider Team, Departamento de Física Fundamental II, and Instituto de Materiales y Nanotecnología, Universidad de La Laguna, La Laguna 38205, Tenerife, Spain

[5]MALTA Consolider Team, DCITIMAC, Facultad de Ciencias, Universidad de Cantabria, 39005 Santander, Spain

**Abstract:** This paper reports an investigation on the phase diagram and compressibility of wolframite-type tungstates by means of x-ray powder diffraction and absorption in a diamond-anvil cell and *ab initio* calculations. The diffraction experiments show that monoclinic wolframite-type $MgWO_4$ suffers at least two phase transitions, the first one being to a triclinic polymorph with a structure similar to that of $CuWO_4$ and $FeMoO_4$-II. The onset of each transition is detected at 17.1 and 31 GPa. In $ZnWO_4$ the onset of the monoclinic-triclinic transition has been also found at 15.1 GPa. These findings are supported by density-functional theory calculations, which predict the occurrence of additional transitions upon further compression. Calculations have been also performed for wolframite-type $MnWO_4$, which is found to have an antiferromagnetic configuration. In addition, x-ray absorption and diffraction experiments as well as calculations reveal details of the local-atomic compression in the studied compounds. In particular, below the transition pressure the $ZnO_6$ and equivalent polyhedra tend to become more regular, whereas the $WO_6$ octahedra remain almost unchanged. Fitting the pressure-volume data we obtained the equation of state for the low-pressure phase of $MgWO_4$ and $ZnWO_4$. These and previous results on $MnWO_4$ and $CdWO_4$ are compared with the calculations, being the compressibility of wolframite-type tungstates systematically discussed. Finally Raman spectroscopy measurements and lattice dynamics calculations are presented for $MgWO_4$.

I. **Introduction**

Materials belonging to the tungstate family ($AWO_4$ with A being a divalent element) have a long history of practical application and have been the object of extensive research. Their optical and luminescence properties have received great attention as these compounds are widely used as scintillating detectors in high-energy particle physics, rare-event searches, and medical diagnosis among other applications [1]. From the fundamental and geophysical standpoints, $AWO_4$ oxides are also interesting compounds [2]. Tunsgtates of large divalent cations (Ca, Ba, Pb, Sr, and Eu) usually crystallize in the tetragonal scheelite structure (space group: $I4_1/a$, Z = 4) and those compounds of small divalent cations (Cd, Zn, Mg, Mn, etc.) can take the wolframite structure (space group: $P2/c$, Z = 2) [3]. In wolframite (shown in Fig. 1), both A and W cations have octahedral oxygen coordination and each octahedron shares two corners with its neighbors [4].

In the last years there has arisen renewed interest in AWO$_4$ compounds and their evolution under pressure. In the case of wolframites, high-pressure (HP) Raman spectroscopy studies have been performed in CdWO$_4$ [5, 6] and ZnWO$_4$ (the mineral sanmartinite) [7, 8]. *Ab initio* calculations have been also carried out to study their structural stability [6, 8]. In ZnWO$_4$ a pressure-induced phase transition has been detected beyond 30 GPa [8] and a subsequent transition suggested near 58 GPa. A monoclinic β-fergusonite-type structure (space group: *C2/c*, Z = 4) has been proposed for the HP phase. In CdWO$_4$ two transitions take place around 20 and 35 GPa, being two coexisting phases found within this pressure range one with tetragonal and another with triclinic symmetry. At 35 GPa CdWO$_4$ transforms to a β-fergusonite-type structure. Under the current status of the research, HP x-ray diffraction studies are needed to further progress on the understanding of the structural properties of wolframites. However, in contrast with scheelite-structured tungstates [1, 9, 10], such studies have been rarely performed in wolframites. Indeed CdWO$_4$, MgWO$_4$, and MnWO$_4$ (the mineral hübnerite) have been studied only up to a pressure lower than 8 GPa [11]. Therefore the information available is limited to the compression of the wolframite phase at low pressures.

Motivated by the above-described facts, we have undertaken a comparative study of the structural properties of CdWO$_4$, MgWO$_4$, MnWO$_4$, and ZnWO$_4$ under compression. In this paper, we report HP angle-dispersive x-ray diffraction (ADXRD) experiments in MgWO$_4$ (ZnWO$_4$) up to nearly 50 (26) GPa as well as x-ray absorption spectroscopy (XAS) measurements in ZnWO$_4$ up to 18 GPa at the Zn K-edge. The obtained results are interpreted with the help of *ab initio* total-energy calculations. In MgWO$_4$, we observe two phase transitions and propose a triclinic structure for the first HP phase. For ZnWO$_4$ an intermediate phase with $P\bar{1}$ symmetry has been found at lower pressure than the proposed phase with *C2/c* symmetry [8], pointing towards the same sequence followed by MgWO$_4$. The equation of state (EOS) for the wolframite phase has been determined. Calculations performed for woframite CdWO$_4$ and MnWO$_4$ are also compared with earlier data. In addition, lattice dynamics properties of MgWO$_4$ have been computed, being the Raman- and infrared-active phonons and their pressure dependences determined. Finally, Raman measurements have been carried out for MgWO$_4$.

II. **Experimental details**

To perform the experiments we used micron-size powder samples with a purity higher than 99% (Aldrich). Their crystal structures at ambient conditions were checked by powder x-ray diffraction using a Seifert XRD 3003 TT diffractometer with Cu K$_\alpha$ monochromatic radiation ($\lambda = 1.5406$ Å). The diffraction patterns of MgWO$_4$ and ZnWO$_4$ correspond to the wolframite-type phase with no indication of additional phases. The unit-cell parameters were $a = 4.689(2)$ Å, $b = 5.675(2)$ Å, $c = 4.928(2)$ Å, and $\beta = 90.75(5)°$ for MgWO$_4$ and $a = 4.680(2)$ Å, $b = 5.712(2)$ Å, $c = 4.933(2)$ Å, and $\beta = 90.30(5)°$ for ZnWO$_4$, being in excellent agreement with previous studies [3, 8, 11].

ADXRD experiments were carried out at beamline I15 in the Diamond Light Source with a monochromatic x-ray beam ($\lambda = 0.6118$ Å), which was focused down to 30 μm x 30 μm using Kirkpatrick-Baez mirrors. In order to pressurize the samples two different diamond-anvil cells (DACs) were used: a membrane-type DAC with diamond-culets of 400 μm and a Boehler-Almax-type DAC with diamond-culets of 280 μm. In the membrane DAC, samples were loaded in 150-μm-holes drilled in Inconel gaskets and in the Boehler-Almax DAC in 100-μm-holes drilled in tungsten gaskets. Silicone oil was used as pressure-transmitting medium [12, 13] and the pressure was measured by means of the ruby fluorescence technique [14]. A pinhole placed before the sample position was used as a clean-up aperture for filtering out the tail of the focused x-ray beam. High-resolution diffraction patterns were recorded using a MAR345 image plate located at a distance of 423 mm from the sample. They were integrated and corrected for distortions using FIT2D. The analysis and indexing of the structures was performed using POWDERCELL, DICVOL, and UNITCELL. One experiment was carried out in $MgWO_4$ and two different runs were performed for $ZnWO_4$.

Dispersive x-ray absorption experiments on $ZnWO_4$ up to 18 GPa were carried out at ODE beamline in the Soleil Synchrotron using a membrane DAC with 400 μm-diameter diamond-culets. Experiments were conducted at the Zn K-edge (9.659 KeV) with a focused polychromatic beam of 50 μm x 50 μm and using metallic Zn as a reference compound for energy calibration. The incident and transmitted beam were systematically measured using a grazing mirror between the sample and the CCD detector in order to avoid the presence of harmonics. The sample, in powder form ($\Delta\mu d \sim 2.1$), was loaded in a 200 μm-diameter hole of an Inconel gasket, together with ruby chips and silicone oil as pressure-transmitting medium.

Unpolarized confocal micro-Raman scattering measurements were performed at room temperature in a double monochromator Jobin-Yvon U1000 equipped with a $N_2$ cooled charge coupled device (CCD) detector, in the backscattering geometry. The 520 and 646 nm lines (wavelengths longer than that of the $MgWO_4$ band gap [15]) of an Coherent Innova Argon-Kripton laser were used at an incident power of 10 mW on the sample, which proved to be low enough to avoid spurious effects caused by the laser induced heating of the sample. This was verified by varying the incident power and observing that neither the Stokes to anti-Stokes intensity ratio nor the frequency of the $A_g$ mode at 916.8 $cm^{-1}$ varied within experimental precision. The laser spot diameter on the sample was 1 μm and the spectral resolution was better that 1 $cm^{-1}$.

## III. Calculations

First-principles total-energy and lattice-dynamics calculations were done within the framework of the density-functional theory (DFT) and the pseudopotential method using the Vienna *ab initio* simulation package (VASP) [16 – 18]. The exchange and correlation energy was initially taken in the local-density approximation (LDA) [19] for $MgWO_4$ and the generalized-gradient approximation (GGA) according to Perdew-Burke-Ernzerhof prescription for $MnWO_4$ [20]. The projector-augmented wave (PAW) scheme [21, 22] was adopted and the semicore $5p$ electrons of W were also explicitly included in the calculations. The set of plane waves used extended up to a kinetic energy cutoff of 520 eV. This large cutoff was required to deal with the O atoms within the PAW scheme to ensure highly converged results. The Monkhorst-Pack (MP) [23] grid used for Brillouin-zone integrations ensured highly converged results (to about 1 meV per formula unit). We use 16, 24, 8 and 6 k-points to perform the geometrical optimizations of $MgWO_4$ with symmetry *P2/c* (wolframite), *P$\bar{1}$*, *C2/c*, and *Cmca*, respectively. For $MnWO_4$ we performed spin density calculations and we found that the antiferromagnetic configuration was the most stable one. For the Brillouin zone integrations we used 16 k-points with an MP grid of (4, 4, 4) for the analyzed structures. At each selected volume, the structures were fully relaxed to their equilibrium configuration through the calculation of the forces on atoms and the stress tensor – see Ref. 24. In the relaxed equilibrium configuration, the forces are less than 0.004 eV/Å and the deviation of the stress tensor from a diagonal hydrostatic form is less than 1 kbar (0.1 GPa). The highly converged results on forces are required for the calculation of the dynamical matrix using the direct force constant approach (or supercell method) [25] .The construction of the dynamical matrix at the Γ point involves separate calculations of the forces in which a fixed displacement from the equilibrium configuration of the atoms within the *primitive* cell is considered. Symmetry aids by reducing the number of such independent distortions, reducing the amount of computational effort in the study of the analyzed structures considered in our work. Diagonalization of the dynamical matrix provides both the frequencies of the normal modes and their polarization vectors, it allows us to identify the irreducible representation and the character of the phonon modes at the zone center. The calculations performed for $ZnWO_4$ and $CdWO_4$ and their results have been described in detail in Refs. 6 and 8.

**IV.    Results and discussion**

**A. X-ray diffraction studies**

A sequence of the evolution with pressure of selected x-ray diffraction patterns of $MgWO_4$ is shown in Fig. 2. All the Bragg reflections can be indexed up to 17.1 GPa according to the wolframite structure. At this pressure additional peaks appear (see Figs. 2 and 3); e.g. the peak located around $2\theta = 7.2º$. As pressure increases these and extra peaks grow in intensity up to 27.4 GPa. This fact suggests the onset of a phase transition at 17.1 GPa from the wolframite structure to a high-pressure one. At 31 GPa an additional peak appears. The appearance of this peak is followed by the appearance of two more at 35.1 GPa and extra ones up to 41.1 GPa. This fact points out to the onset of a second phase transition at 31 GPa. Both transitions are reversible as it can be seen on the diffraction pattern collected at 0.2 GPa after pressure release. This pattern is similar to the one obtained at 0.6 GPa before compression, being the main difference the peak broadening. In $ZnWO_4$ similar changes in the diffraction patterns than in $MgWO_4$ have been found (see Fig. 4), with a phase transition onset at about 15.1 GPa, pointing to a similar phase transition than in $MgWO_4$.

Regarding the crystalline structure of the post-wolframite phase, in both compounds we can tell that the diffraction patterns suggest a reduction of the crystal symmetry after the phase transition. Given the small number of reflections and their broad nature we have not been able to perform a Rietvelt refinement for either of the HP phases. As a consequence of it, the structure of the second HP phase of MgWO$_4$ remains yet undetermined. However, an indexing of the diffraction patterns measured for MgWO$_4$ at 27.4 GPa indicated that its first HP phase has a triclinic $P\bar{1}$ symmetry with $Z = 2$, in agreement with previous Raman studies from intermediate wolframite structures in related compounds [26]. In particular, at 27.4 GPa we obtain the following unit-cell parameters for the potential triclinic polymorph of MgWO$_4$: $a = 4.49(1)$ Å, $b = 5.43(1)$ Å, $c = 4.86(1)$ Å, $\alpha = 92.3(5)°$, $\beta = 87.0(5)°$, and $\gamma = 90.5(5)°$. Therefore, this phase resembles the structure of CuWO$_4$ [27] and FeMoO$_4$-II [28]. This structure is a distorted version of wolframite, which is topologically related to it. Note that the triclinic space group is a subgroup of the monoclinic space group and consequently the proposed transition appears to be continuous. This is consistent with our finding that the coordination of the Mg and W atoms does not change at the transition. Comparing both structures in Fig. 1, it can be seen that the phase transition mainly consists on a movement of the oxygen and A atoms, implying a rotation of the octahedron and a distortion of it. This distortion is enough to reduce the symmetry to triclinic. Basically, the proposed transition involves a tilting and distortion of the MgO$_6$ octahedra, which produces a local symmetry breakdown into the triclinic symmetry. In the triclinic phase, the Mg atoms have a more irregular coordination compared with wolframite. As we will show later, a similar structure is predicted by our *ab initio* calculations for the HP phase of MgWO$_4$. Another evidence supporting the occurrence of a phase transition at 17.1 GPa is given in Fig. 5, where the full-width at half maximum (FWHM) of three Bragg peaks is represented as a function of pressure. A steep increase in their width is seen at that pressure, coinciding with the transition onset. This fact can be explained by comparing the diffraction patterns of both structures. As it is characteristic of second-order transitions, the wolframite peaks can be also accounted for with the new triclinic structure. However, the triclinic distortion gives rise to a contribution from extra peaks that split from the original wolframite peaks explaining the observed peak broadening.

For $ZnWO_4$, our x-ray diffraction experiments suggest that the HP phase has the same triclinic structure as HP $MgWO_4$. This is apparently in contradiction with previous Raman-spectroscopy experiments [8], which reported the onset of a phase transition to a monoclinic β-fergusonite-type phase (*C/2c*) beyond 30.6 GPa. However, the 15.1 GPa transition pressure is very close to the pressure were a domain formation was observed in single-crystalline $ZnWO_4$ together with a relative change of the Raman peak intensities [8]. A possible explanation for this discrepancy can be the use of different pressure-transmitting media (Ar in Raman experiments and silicone oil in present experiments). According with *ab initio* calculations the triclinic $P\bar{1}$ structure is energetically competitive with the wolframite (see Fig. 4 of Ref. 8) and β-fergusonite-type structures. Therefore the non-hydrostatic stresses induced by the use of silicone oil as pressure medium could favor the transition to a metaestable triclinic structure. In our ADXRD experiments the maximum pressure reached is below the onset of the transition to the *C2/c* structure, which prevents us to make any conclusions about the structure of this phase. This explanation is also consistent with the rich polymorphism observed in other orthotungstates under compression, e.g. $BaWO_4$ and $PbWO_4$ [29] and the fact that kinetics has an important effect in their pressure-driven phase transitions. Note that the triclinic structure was also observed in $CdWO_4$ as an intermediate phase between the low-pressure wolframite structure and the high-pressure β-fergusonite phase [6]. Therefore, in addition to MgWO$_4$ and CdWO$_4$, the triclinic phase can be also pressure-induced in ZnWO$_4$ several GPa below the β-fergusonite phase.

From the analysis of our x-ray diffraction data, we extracted the pressure dependence of the lattice parameters for both wolframite MgWO$_4$ and ZnWO$_4$. These results are summarized in Fig. 6. This behaviour is in agreement with that reported by Macavei *et al.* from single crystal studies up to 8 GPa [11]. In Fig. 7 we report the evolution of the unit-cell volumes of MgWO$_4$ and ZnWO$_4$ obtained in our experiments as well as the volume data obtained previously [11] for MgWO$_4$, MnWO$_4$ and CdWO$_4$. In the case of MgWO$_4$ our data are in good agreement with single-crystal diffraction data within the pressure range covered by these measurements [11]. We have analyzed the volume changes using a third-order Birch-Murnaghan EOS [30]. The obtained EOS parameters for MgWO$_4$ [ZnWO$_4$] are: $V_0$ = 131.1(3) Å$^3$ [132.9(5)], $B_0$ = 160(13) GPa [145(6)], and $B_0$'= 4.5(3) [6.6(9)], these parameters being the zero-pressure volume, bulk modulus, and its pressure derivative, respectively.

**B. X-ray absorption measurements**

Fig. 8 shows a selection of extended x-ray absorption fine structure (EXAFS) signals measured in ZnWO$_4$ at different pressures. The EXAFS signal remains stable under pressure, showing a slight displacement towards higher k values as pressure is increased due to the shortening of the Zn-O distances. In order to analyze these results, we obtained the pair-pseudo-distribution function (PPDF) by Fourier Transformation of the EXAFS oscillations in the [1.2, 9] k range, using a Bessel-type (Z = 4) apodization window (see Fig. 9). The main peak is associated to the first oxygen neighbor shell. The six distances in the ZnO$_6$ octahedron can be grouped in one short (x4) and one long distance (x2). The position at about 1.42 Å of this broad peak - 0.8 Å FWHM - slowly changes with pressure pointing towards smaller R values with almost identical shape and FWHM. The fit to the main peak (see Fig.9) has been made with the WinXAS program and calculating the corresponding phases and amplitudes with the *ab initio* FEFF 8.6 code [31] for multiple-scattering x-ray absorption fine structure.

From the analysis of the EXAFS data we obtained the evolution with pressure of the Zn-O distances. These results are shown in Fig. 10 together with those obtained from our diffraction measurements in the wolframite phase -considering no variation of the Wychoff positions with pressure- and *ab initio* calculations. In particular, we found that the different methods give a qualitatively similar behavior. In both wolframites there is a noticeable difference in the compressibility shown under pressure between the W-O and the A-O distances, being the compressibility of the later at least three times the one of the former. This fact means that it can be considered that $WO_6$ octahedra behave as rigid units while $MgO_6$ ($ZnO_6$) octahedra account for most of the crystal compressibility. If we compare the evolution for the distances calculated and those obtained through the ADXRD experiment it can be pointed out that for W-O bonds in $ZnWO_4$, the assumption that Wychoff positions do not change with pressure seems to be a good approximation. However, this simplification does not work so well for the Zn-O distances. In the case of the $MgWO_4$ the same approximation appears to work well below 6 GPa, since bond distances obtained assuming no changes in the atomic positions are similar to those obtained when a structural refinement was performed at different pressures [11]. The only differences are seen for the longest bond that shows to be more compressible than in the *ab initio* calculations and our ADXRD results.

Regarding the calculations, even though the theoretical Mg-O distances differ in absolute value from the experimental ones, the bond compressibilities found are in good accordance with ADXRD results. In the case of the W-O distances of $MgWO_4$, the agreement between the experiments and the calculations is better for the compressibility of the longest W-O distance than for the shortest ones. The largest Zn-O distance obtained through EXAFS is in very good agreement with the one obtained by the theoretical calculations as well as with the high compressibility found in the A-O bond found by Macavei et al. [11] in several wolframites. This points out that considering the Wychoff positions as constant is not the best approximation meaning that the $ZnO_6$ octahedra are modified under pressure. Regarding the shortest distance obtained from EXAFS, it is slightly underestimated and almost does not change with pressure. This can be caused by the fact that this distance is an average of four different distances, having apparently little influence in the position of the FT peak.

We would like to mention now that we have not observed any detectable change in the X-ray Absorption Near Edge Structure (XANES) signal in $ZnWO_4$ under high pressure up to 17.4 GPa. Since XANES originates from multiple scattering of photoelectrons, it is very sensitive to changes in symmetry around the absorbing atom. Thus the unchanged XANES signal is in agreement with the fact that $ZnWO_4$ remains in the wolframite structure up to nearly 17 GPa.

**C. Raman measurements**

In contrast with $CdWO_4$ and $ZnWO_4$, Raman spectroscopy data have not been reported up to now in $MgWO_4$. According to Group Theory, the wolframite structure presents at the $\Gamma$ point 18 Raman active modes ($8A_g + 10B_g$) and 18 infrared ones ($8A_u + 10B_u$). In this kind of compounds [8, 26], due to the difference in valence and weight between $W^{6+}$ atoms and $A^{2+}$, the lattice dynamics can be understood assuming two kinds of modes. Six of them related with motions inside the covalently bounded $WO_6$ octahedra called internal and other modes originated as motions of the A cations against the $WO_6$ units named external. The internal modes are at higher frequencies than the external ones and usually there is a phonon gap between them [6, 8]. According to our calculations the internal modes of the $MgWO_4$ are the $A_g$ modes at 420.4, 551.6, 713.2 and 916.8 cm$^{-1}$ and the $B_g$ modes at 683.9 and 808.5 cm$^{-1}$.

Two Raman spectra measured at ambient conditions for $MgWO_4$ are presented in Fig. 11 with the inset being measured with the 646 nm wavelength laser up to 600 cm$^{-1}$ and showing one mode (215 cm$^{-1}$) not detected with the other wavelength (520 nm). Due to the Raman set-up used, we have not been able to get closer to the Rayleigh line more than 200 cm$^{-1}$. Nevertheless we have been able to observe 14 out of the 18 Raman modes being their frequencies (ω) in good agreement with our calculations (see Table I). About the remaining modes, according to theory, three of them are below 200 cm$^{-1}$, while one may not be detected due to the overlap of the $B_g$ and $A_g$ modes calculated to be at 405.16 and 411.27 cm$^{-1}$, respectively. In table, I the theoretical Grüneisen parameters are also included ($\gamma = \frac{B_0}{\omega}\frac{\partial \omega}{\partial P}$). The agreement obtained between experiments and calculations at ambient pressure validates our theoretical predictions for high pressure. In Table II the calculated infrared active modes as well as their pressure coefficients and Grüneisen parameters are shown for completeness. There are three infrared modes that soften upon compression.

**D. Theoretical Results**

In order to help with the interpretation of our experimental results, *ab initio* total-energy and lattice-dynamics calculations were performed for $MgWO_4$. Similar calculations were performed for $MnWO_4$ and previously for $CdWO_4$ [6] and $ZnWO_4$ [8]. These results will be systematically compared with the bulk of experimental results available for wolframite tungstates. For $MgWO_4$, along with the wolframite structure, we have considered other structures on account of their observation or postulation in previous HP works for related compounds: $CuWO_4$-type ($P\bar{1}$), orthorhombic-disorder wolframites (*A2$_1$am, Pmma, Pbam, Pbcm, Pbcn, Pbnm, Cmcm*, and *Cmca* from LaTaO$_4$, CaRhIn$_4$, YbCoB$_4$, BaUO$_4$, NaScCl$_4$, MgSeO$_4$, CaSO$_4$, and PbWO$_4$, respectively), M-fergusonite (*I2/a*), M'-fergusonite (*P2$_1$/c*), β-fergusonite (*C2/c*), monoclinic-disorder wolframites (*Pc, P2/m, C2/m,* and *P2$_1$/n* from KAuCl$_4$, CrAuTe$_4$, CrPS$_4$, and AgMnO$_4$, respectively), and scheelite (*I4$_1$/a*). Figure 12 shows the energy vs. volume curves for the most competitive structures. The common tangent construction enables to deduce the transition and equilibrium pressures [32]. In the figure we only reported those structures that play a relevant role in the HP structural behavior of MgWO$_4$. According to the calculations wolframite (*P2/c*) is the most stable structure at ambient pressure. The calculated lattice parameters for this structure are: $a$ = 4.6332 Å, $b$ = 5.6197 Å, $c$ = 4.8831, and β = 90.36. These parameters are slightly smaller than the experimental values. The differences are within the typical reported systematic errors in DFT-LDA calculations. The calculated EOS of wolframite MgWO$_4$ is given by the following parameters $V_0$ = 127.2 Å$^3$, $B_0$=161 GPa, and $B_0$' = 4.2. They were obtained using the Birch-Murnaghan EOS [30]. Above 35.4 GPa the calculations predict that the most stable structure would be a β-fergusonite one (*C2/c*). Its lattice parameters and Wychoff positions are in Table III together with the orthorhombic structure (*Cmca*) predicted to become stable beyond 48 GPa as well as the structural information about wolframite. The proposed transition to the β-fergusonite structure is possible given the evidence available in CdWO$_4$ and ZnWO$_4$ [6, 8]. The stabilization of the orthorhombic structure involves a coordination increase for Mg from 6 to 10. The coordination of W remains unaltered. The appearance of this structure is in agreement with the HP systematic developed by Bastide [1] and with the structural sequence found in other tungstates [33]. The confirmation of the existence of the β-fergusonite and *Cmca* structures is waiting for future experiments.

Regarding the triclinic $P\bar{1}$ structure, the calculations also predict that in the pressure range where it was experimentally observed, it is energetically competitive with wolframite. This means that possible uniaxial stresses induced by the use of a pressure medium like silicone oil could be enough to induce a phase transition from the wolframite structure to the triclinic one. This fact could probably be the cause of the finding of the triclinic phase instead of the β-fergusonite phase beyond 17.1 GPa. In Table III we report the calculated structural information for the $P\bar{1}$ structure at 29 GPa. The agreement with the experimental values is good.

As we mentioned above, calculations have been performed for the low-pressure phase of $MnWO_4$. Since this compound could be magnetic due to the presence of the Mn cation, we considered different magnetic configurations. We found the low-pressure phase of $MnWO_4$ to have a wolframite structure with an antiferromagnetic configuration. In this configuration, Mn has a magnetic moment of 4.319 μB while in the ferromagnetic configuration is 4.356 μB. The structural information for this phase is summarized in Table IV. The agreement with the literature [11] is good. For this phase we also calculated its EOS obtaining the following parameters: $V_0$ = 157.4 Å$^3$, $B_0$ = 125 GPa, and $B_0$' = 4.3. The bulk modulus agrees well with the previously reported experimental value [11].

To conclude we would like to compare the different compressibilities reported for $MgWO_4$, $MnWO_4$, $CdWO_4$ and $ZnWO_4$. A summary of their bulk moduli and unit-cell volume are given in Table V. It is straightforward to see there, that there is an inverse correlation between the volume and the bulk modulus. As expected the densest compounds are the less compressible. It can be also seen that our experimental values agree with previous values within 10%. The same conclusion can be extracted for our calculations.

Some insight on the compressibility of wolframites can be extracted from their polyhedral compressibility. As we discussed above, W-O bonds are much more rigid than A-O bonds. Then, the compression of wolframite can be attributed dominantly to changes in the A-O bond lengths. The same trend has been found in scheelite-structured orthotungstates and orthomolybdates [33, 34], leading to a phenomenological rule that correlates the bulk modulus with the inverse of the A-O distance. If the same rule is applied to wolframite, we estimate the bulk modulus given in the right hand-side column of Table V. As can be seen there, the estimated values reproduce the tendency followed by the bulk compressibility of wolframites, but slightly underestimates the bulk modulus. However, we think, the phenomenological rule [33] can be used to roughly estimate $B_0$ of unstudied wolframites like $FeWO_4$ (134 GPa), $NiWO_4$ (141 GPa), and $CoWO_4$ (137 GPa). Finally, the different compressibility of $AO_6$ and $WO_6$ octahedra also explains that the compaction of wolframite is more important in the *b*-axis than in other axes. Basically, it is related to the different linking of octahedral units along different crystallographical directions.

**V. Conclusions**

By means of x-ray diffraction and absorption experiments and *ab initio* calculations we have studied the structural properties of different wolframite compounds up to 50 GPa. For $MgWO_4$ we have found evidences of two pressure-induced phase transitions at 17.1 and 31 GPa. A phase transition was detected in $ZnWO_4$ at 15.1 GPa, a lower pressure than that of the previously reported one [8] (30 GPa). Apparently, the first transition involves a symmetry-reduction in the crystal from monoclinic to triclinic. The proposed-structure for the HP phase is similar to that of $CuWO_4$. We also report an EOS for $MgWO_4$, $MnWO_4$, $ZnWO_4$, and $CdWO_4$. Information on the bond compressibility is reported too and related with the bulk compressibility of wolframite. Raman measurements and lattice-dynamics calculations are also presented for MgWO$_4$. Most of its modes have been identified and their Grüneisen parameters reported. Finally, different magnetic configurations have been considered for wolframite MnWO$_4$, being found that it is antiferromagnetic.

**Acknowledgments**

ADXRD experiments carried out at the Diamond Light Source (I15 beamline, proposal No. 683). The authors thank A. Kleppe for technical support. Research financed by Spanish MEC (MAT2007-65990-C03-01/03 and CSD-2007-00045) as well as Mexican CONACyT (J-59853-F and J-83247-F). J. R.-F. (R. L.-P) thanks the MEC support through FPI (FPU) program. We thank SOLEIL for granting us beamtime and the ODE staff for technical assistance.

**References**

[1] D. Errandonea and F. J. Manjon, Progress in Materials Science **53**, 711 (2008) and references therein.

[2] D. Errandonea, D. Martínez-García, R. Lacomba-Perales, J. Ruiz-Fuertes, and A. Segura, Appl. Phys. Letters **89**, 091913 (2006).

[3] A. W. Sleight, Acta Cryst. B **28**, 2899 (1972).

[4] R. O. Keeling, Acta Cryst. **10**, 209 (1957).

[5] A. Jayaraman, S. Y. Wang, and S. K. Sharma, Curr. Sci. **69**, 44 (1995).

[6] R. Lacomba-Perales, D. Errandonea, D. Martinez-Garcia, P. Rodríguez-Hernández, S. Radescu, A. Mujica, A. Muñoz, J. C. Chervin, and A. Polian, Phys. Rev. B **79**, 094105 (2009).

[7] A. Perakis, E. Sarantapoulou, and C. Raptis, High Press. Res. **18**, 181 (2000).

[8] D. Errandonea, F. J. Manjon, N. Garro, P. Rodriguez-Hernandez, S. Radescu, A. Mujica, A. Muñoz, and C. Y. Tu, Phys. Rev. B **78**, 054116 (2008).

[9] D. Errandonea, M. Somayazulu, and D. Häusermann, phys. stat. sol. (b) **235**, 162 (2003).

[10] D. Errandonea, phys. stat. sol. (b) **242**, R125 (2005).

[11] J. Macavei and H. Schulz, Z. Kristallog. **207**, 193 (1993).

[12] Y. Shen, R. S. Kumar, M. Pravica, and M. F. Nicol, Rev. Sci. Insrum. **75**, 4450 (2004).

[13] D. Errandonea, Y. Meng, M. Somayazulu, and D. Häusermann, Physica B **355**, 116 (2005)

[14] H. K. Mao, J. Xu, and P. M. Bell, J. Geophys. Res. **91**, 4673 (1986).

[15] R. Lacomba-Perales, J. Ruiz-Fuertes, D. Errandonea, D. Martínez-García, and A. Segura, EPL **83**, 37002 (2008).

[16] G. Kresse and J. Hafner, Phys. Rev. B **47**, 558 (1993).

[17] G. Kresse and J. Hafner, Phys. Rev. B **49**, 14251 (1994).

[18] Kresse and J. Furthmüller, Phys. Rev. B **54**, 11169 (1996).

[19] J. P. Perdew and A. Zunger, Phys. Rev. B **23**, 5048 (1981).

[20] J. Perdew, K. Burke, and M. Ernzerhof, Phys. Rev. Lett. **78**, 1396 (1997).

[21] P.E. Blochl, Phys. Rev. B **50**, 17953 (1994).

[22] G. Kresse and D. Joubert, Phys. Rev. B **59**, 1758 (1999).

[23] H.J. Monkhorst and J.D. Pack, Phys. Rev. B **13**, 5188 (1976).

[24] F. J. Manjón, D. Errandonea, N. Garro, J. Pellicer-Porres, P. Rodríguez-Hernández, S. Radescu, J. López-Solano, A. Mujica, and A. Muñoz, Phys. Rev. B **74**, 144111 (2006).

[25] K. Parlinski, computer code PHONON. See: http://wolf.ifj.edu.pl/phonon.

[26] A. Jayaraman, G. A. Kourouklis, L. G. Van Uitert, W. H. Grodkiewicz and R. G. Maines. Physica A **156**, 325 (1988).

[27] L. Kihlborg and E. Gebert. Acta Cryst. B**26**, 1020 (1970).

[28] A. W. Sleigh, B. L. Chamberland, and J. F. Weiher. Inorg. Chem. **7** (6), 1093 (1968).

[29] R. Lacomba-Perales, D. Martínez-García, D. Errandonea, Y. Le Godec, J. Philippe and G. Morard, High Pressure Research **29**, 76 (2009).

[30] F. Birch, J. Geophys. Res. **83**, 1257 (1978).

[31] J. J. Rehr, R. C. Albers, and S. I. Zabinsky, Phys. Rev. Lett. **69**, 3397 (1992).

[32] A. Mujica, A. Rubio, A. Muñoz, and R. J. Needs, Rev. Mod. Phys. **75**, 863 (2003).

[33] D. Errandonea, J. Pellicer-Porres, F. J. Manjón, A. Segura, Ch. Ferrer-Roca, R. S. Kumar, O. Tschauner, P. Rodríguez-Hernández, J. López-Solano, S. Radescu, A. Mujica, A. Muñoz, and G. Aquilanti, Phys. Rev. B **72**, 174106 (2005).

[34] D. Errandoneaa, R. S. Kumar, X. Ma, and C.Y. Tu, J. Sol. State Chem. **181**, 355 (2008).

[35] Y. V. Pisarevskii, I. M. Silvestrova, R. Voszka, A. Peter, I. Foldvari, and J. Janszky, phys. stat. sol. (a) **107**, 161 (1988).

[36] D. M. Trots, A. Senyshyn, L. Vasylechko, R. Niewa, T. Vad, V. B. Mikhailik, and H. Kraus. J. Phys.: Condens. Matter **21**, 325402 (2009).

**Table I**. *Ab initio* calculated and experimental zero pressure frequencies as well as calculated pressure coefficients, and Grüneisen parameters of the Raman modes in wolframite $MgWO_4$.

| Mode | Theory | | | Experiment |
|---|---|---|---|---|
| | $\omega$ ($cm^{-1}$) | $d\omega/dP$ ($cm^{-1}$ GPa) | $\gamma$ | $\omega$ ($cm^{-1}$) |
|---|---|---|---|---|
| $B_g$ | 104.34 | 0.80 | 1.38 | -- |
| $A_g$ | 152.11 | 0.24 | 0.30 | -- |
| $B_g$ | 184.60 | 0.44 | 0.45 | -- |
| $B_g$ | 215.29 | 0.62 | 0.54 | 215.0 |
| $B_g$ | 267.66 | 1.01 | 0.71 | 266.7 |
| $A_g$ | 286.98 | 0.51 | 0.34 | 277.1 |
| $A_g$ | 301.55 | 1.93 | 1.17 | 294.1 |
| $B_g$ | 308.83 | 1.79 | 1.05 | 313.9 |
| $A_g$ | 361.77 | 4.20 | 2.03 | 351.9 |
| $B_g$ | 372.27 | 3.90 | 1.85 | 384.8 |
| $B_g$ | 405.16 | 5.42 | 2.31 | -- |
| $A_g$ | 411.27 | 1.67 | 0.75 | 420.4 |
| $B_g$ | 523.38 | 3.31 | 1.15 | 518.1 |
| $A_g$ | 560.94 | 3.33 | 1.08 | 551.6 |
| $B_g$ | 683.23 | 4.34 | 1.15 | 683.9 |
| $A_g$ | 720.73 | 3.34 | 0.85 | 713.2 |
| $B_g$ | 809.79 | 4.14 | 0.94 | 808.5 |
| $A_g$ | 912.50 | 3.61 | 0.73 | 916.8 |

**Table II**. *Ab initio* calculated zero pressure frequencies, pressure coefficients, and Grüneisen parameters of the Infrared modes in wolframite $MgWO_4$.

| Mode | $\omega$ (cm$^{-1}$) | $d\omega/dP$ (cm$^{-1}$/GPa) | $\gamma$ |
|---|---|---|---|
| $B_u$ | 0 | - | - |
| $A_u$ | 1.77 | - | - |
| $B_u$ | 1.7 | - | - |
| $B_u$ | 193.74 | -0.23 | -0.20 |
| $B_u$ | 241.54 | -0.77 | -0.62 |
| $B_u$ | 268.40 | -0.76 | -0.56 |
| $A_u$ | 278.94 | 1.02 | 0.69 |
| $B_u$ | 315.73 | 3.62 | 2.01 |
| $A_u$ | 331.98 | 0.30 | 0.17 |
| $A_u$ | 341.78 | 1.58 | 0.85 |
| $B_u$ | 366.94 | 4.97 | 2.33 |
| $A_u$ | 445.89 | 4.30 | 1.70 |
| $B_u$ | 482.75 | 5.18 | 1.89 |
| $A_u$ | 519.05 | 4.39 | 1.51 |
| $B_u$ | 575.66 | 3.38 | 1.07 |
| $A_u$ | 685.70 | 3.47 | 0.92 |
| $B_u$ | 777.94 | 3.30 | 0.78 |
| $A_u$ | 869.40 | 3.22 | 0.68 |

**Table III**. Structural information of the calculated structures for $MgWO_4$.

| | $P2/c$ (1 atm) | $P\bar{1}$ (29.0 GPa) | $C2/c$ (35.4 GPa) | $Cmca$ (48 GPa) |
|---|---|---|---|---|
| $a$ | 4.6332 | 4.414 | 6.123 | 7.131 |
| $b$ | 5.6197 | 5.3632 | 9.183 | 10.375 |
| $c$ | 4.8831 | 4.689 | 4.744 | 4.974 |
| $\alpha$ | 90.00 | 90.00 | 90.00 | 90.00 |
| $\beta$ | 90.36 | 89.34 | 131.10 | 90.00 |
| $\gamma$ | 90.00 | 90.00 | 90.00 | 90.00 |
| Mg | 2f (0.5, 0.6691, 0.25) | 2i (0.5, 0.6729, 0.25) | 4e (0, 0.3272, 0.25) | 8e (0.25, 0.3585, 0.25) |
| W | 2e (0, 0.1823, 0.25) | 2i (0, 0.1969, 0.25) | 4e (0, 0.8801, 0.25) | 8f (0, 0.1064, 0.2599) |
| $O_1$ | 4g (0.2171, 0.8955, 0.4360) | 2i (0.2577, 0.5982, 0.5761) | 8f (0.2887, 0.4542, 0.3852) | 8e (0.25, 0.1743, 0.25) |
| $O_2$ | 4g (0.2547, 0.3772, 0.4005) | 2i (0.2377, 0.0939, 0.5668) | 8f (0.7186, 0.2077, 0.2583) | 8f (0, 0.2770, 0.0502) |
| $O_3$ | | 2i (0.2577, 0.4018, 0.0761) | | 8d (0.3375, 0, 0) |
| $O_4$ | | 2i (0.2377, 0.9061, 0.0669) | | 8f (0, 0.4208, 0.3764) |

**Table IV**. Unit-cell parameters and Wychoff positions of MnWO$_4$.

| a = 4.7958 Å, b = 5.8007 Å, c = 5.0254 Å, β = 90.972º | | | | |
|---|---|---|---|---|
| Atom | Site | x | y | z |
| Mn | 2f | 0.5 | 0.6734 | 0.25 |
| W | 2e | 0 | 0.1760 | 0.25 |
| O1 | 4g | 0.2137 | 0.1068 | 0.0636 |
| O2 | 4g | 0.2561 | 0.3726 | 0.3977 |

**Table V**. Bulk modulus and unit-cell volume for different wolframites obtained by ab initio calculations, experiments, and phenomenological estimates. [a]Ref. 11, [b]Ref. 35, and [c]Ref. 36.

| | Volume (Å$^3$) | | | Bulk modulus (GPa) | | |
|---|---|---|---|---|---|---|
| Compound | Experimental values | Ab initio calculations | Previous Experiments | Our Experiments | Ab initio calculations | Phenomenological estimates |
| MgWO$_4$ | 131.1 –132[a] | 127.2 | 144[a] | 160(13) | 161 | 141 |
| ZnWO$_4$ | 132.9 | 137.4 | 153[b] – 161[c] | 145(6) | 140 | 140 |
| MnWO$_4$ | 138.8[a] | 139.8 | 131[a] | | 125 | 125 |
| CdWO$_4$ | 149.3[a] | 157.4 | 136[a] | | 125 | 120 |

**Figures Captions**

**Figure 1 (color online)**. Schematic view of the wolframite structure with a detail of both octahedra (left) and of the $P\bar{1}$ triclinic structure (right).

**Figure 2**. Selection of MgWO$_4$ ADXRD patterns at different pressures from 0.6 GPa to 45.5 GPa.

**Figure 3**. MgWO$_4$ peaks evolution with pressure. Circles: wolframite, squares: triclinic phase, triangles: second HP phase. The vertical lines indicate transition pressures.

**Figure 4.** Selection of ZnWO$_4$ ADXRD patterns at different pressures from 0.7 GPa to 25.9 GPa.

**Figure 5.** Full-width at half maximum evolution with pressure for three independent Bragg peaks in MgWO$_4$.

**Figure 6**. Lattice parameters evolution of MgWO$_4$ and ZnWO$_4$ on the monoclinic structure. Solid symbols: this work. Empty symbols: Ref. 11. Lines: *ab initio* calculations.

**Figure 7**. EOS for the MnWO$_4$, MgWO$_4$, CdWO$_4$ and ZnWO$_4$ wolframites. The empty (filled) circles correspond to Macavei *et al.* [11] (present work), the dotted (solid) lines correspond to *ab initio* calculations (fitted EOS).

**Figure 8**. ZnWO$_4$ EXAFS spectra for several pressures from 2.3 to 17.4 GPa. The spectrum marked as 1.9d corresponds to the one obtained on pressure release.

**Figure 9 (color online)**. Pair-pseudo-distribution function (PPDF) obtained by Fourier transformation of the EXAFS signal at 3.2 GPa. The dot line corresponds to the calculation.

**Figure 10**. A-O and W-O distances for the MgWO$_4$ and ZnWO$_4$ wolframites. The empty circles are the distances obtained by Macavei *et al.* [11]. The solid ones are the distances obtained in the present work through the ADXRD study. The triangles

correspond to the distances obtained with EXAFS. The dot lines are the *ab initio* calculations.

**Figure 11 (color online)**. Raman spectrum of the $MgWO_4$ at ambient conditions. The theoretical and experimental positions of the phonons are also shown.

**Figure 12 (color online)**. *Ab initio* energy vs volume curves for the most stable structures found for the $MgWO_4$ calculation.

**Figure 1**

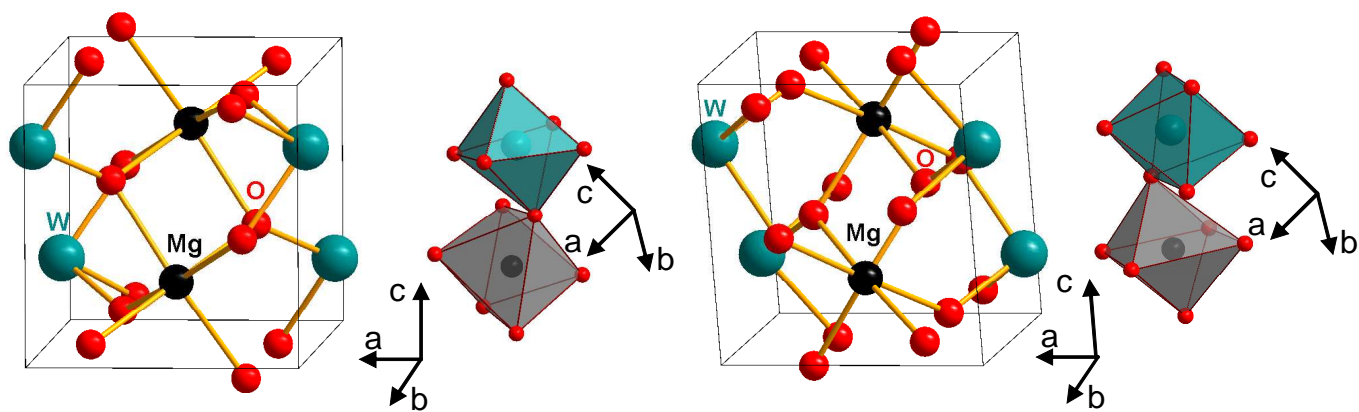

**Figure 2**

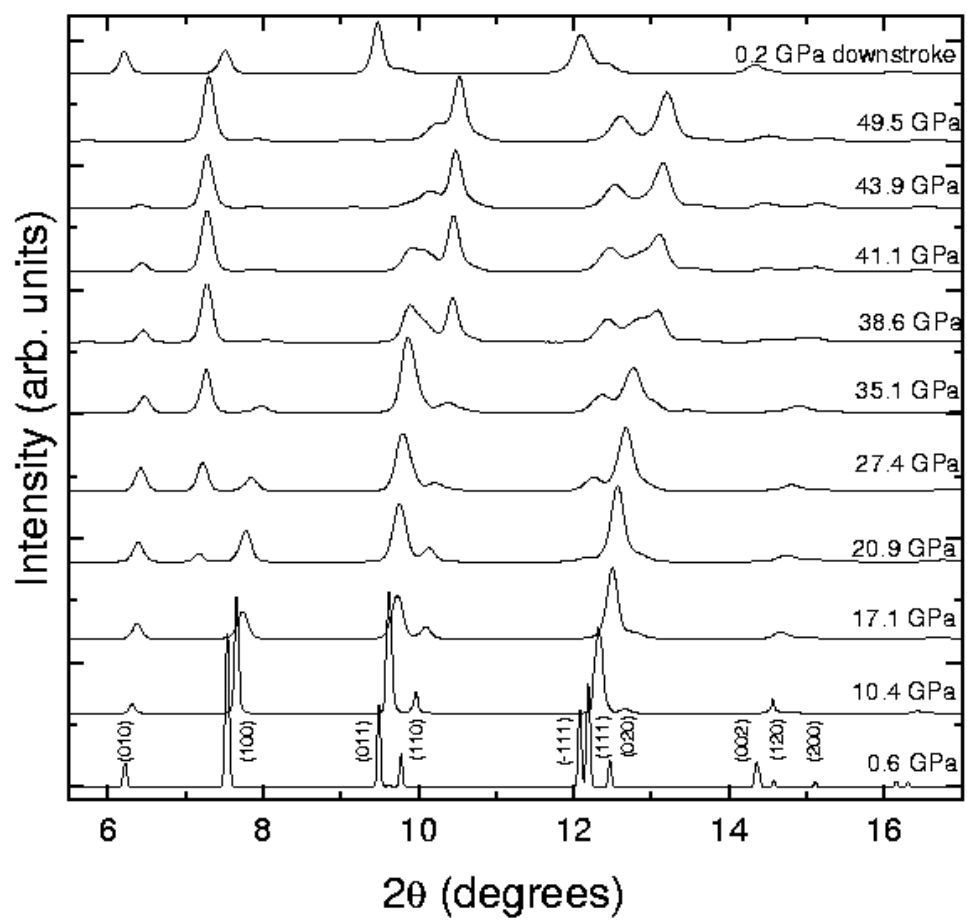

**Figure 3**

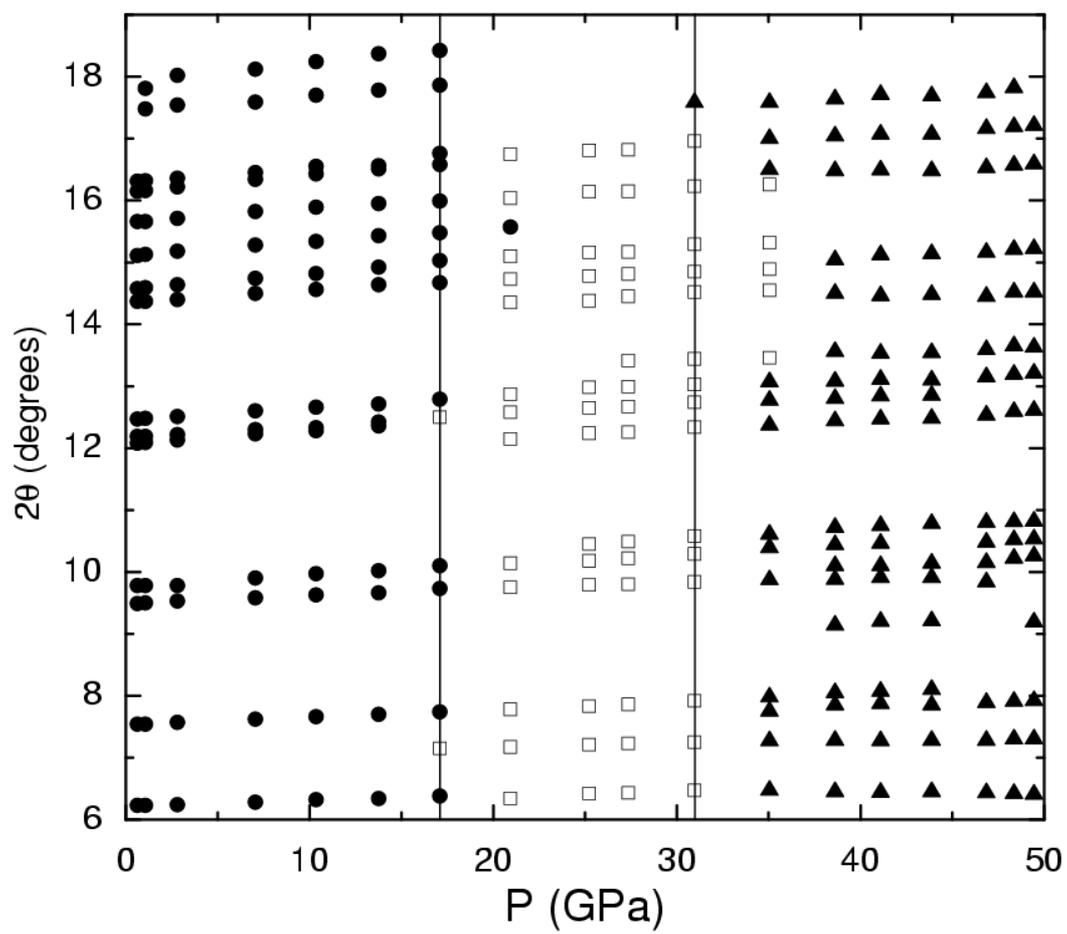

**Figure 4**

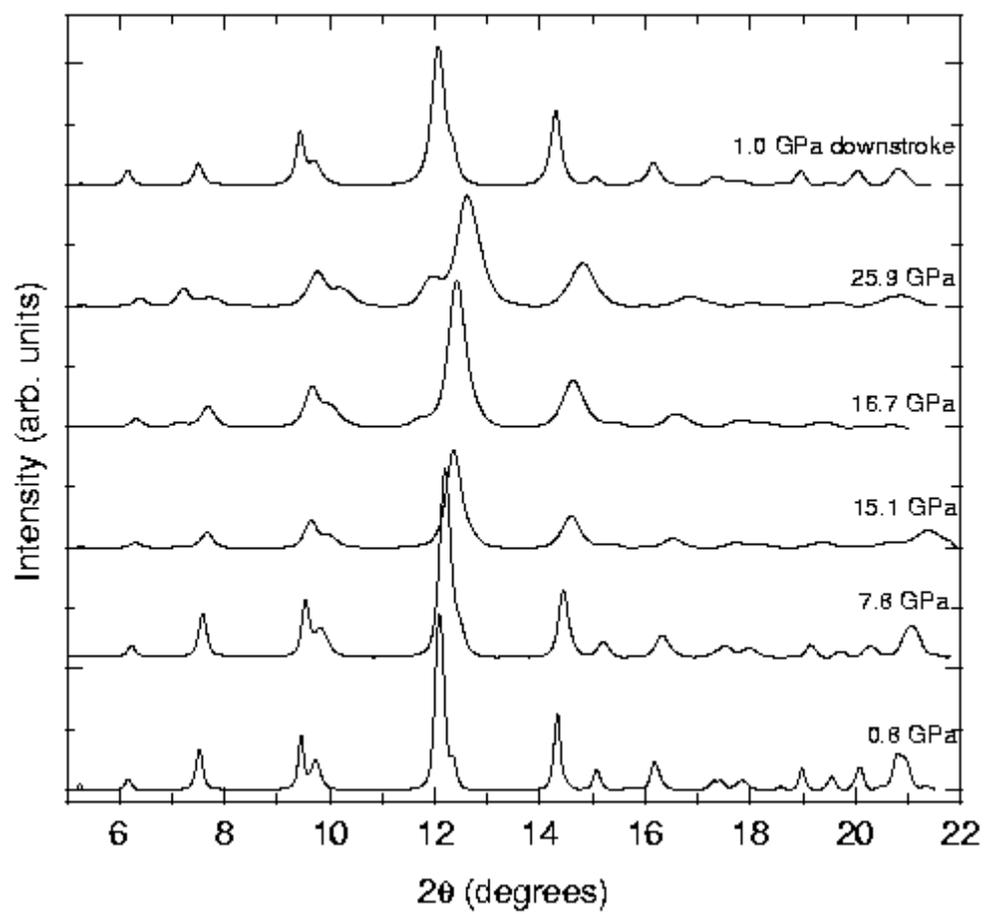

**Figure 5**

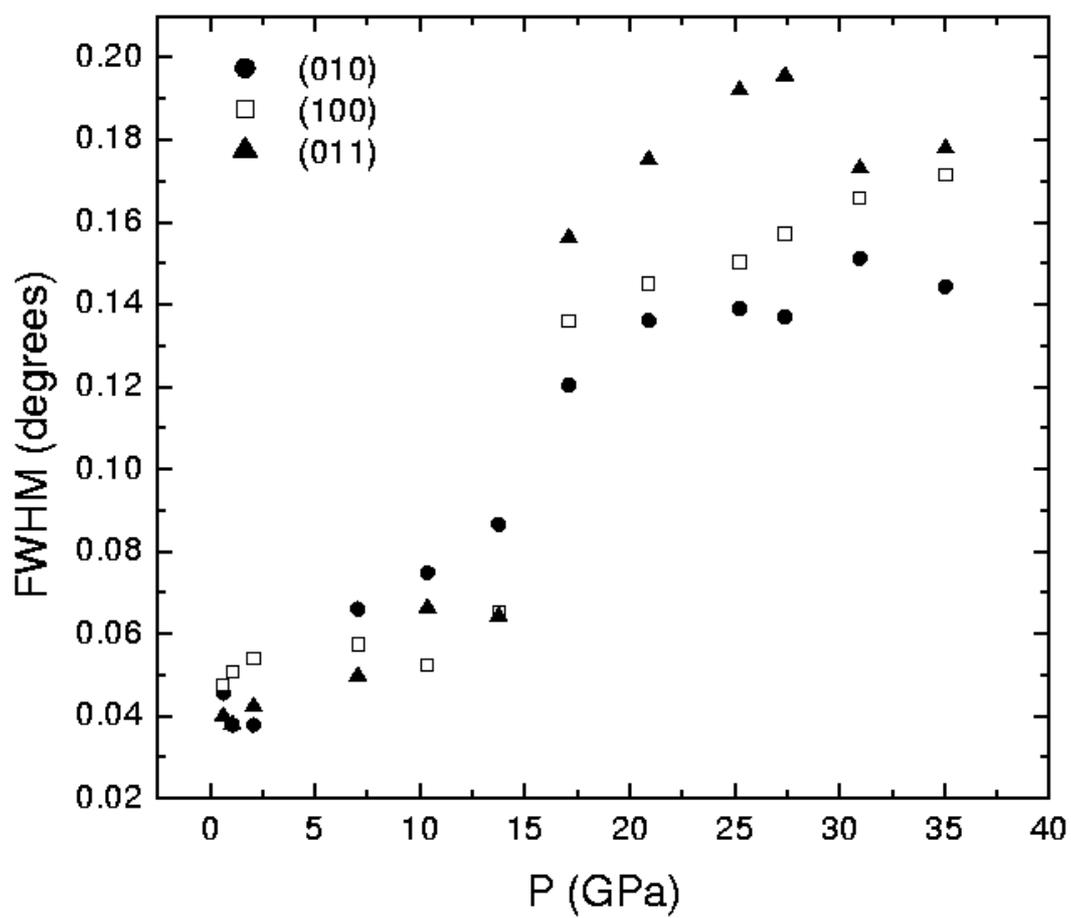

**Figure 6**

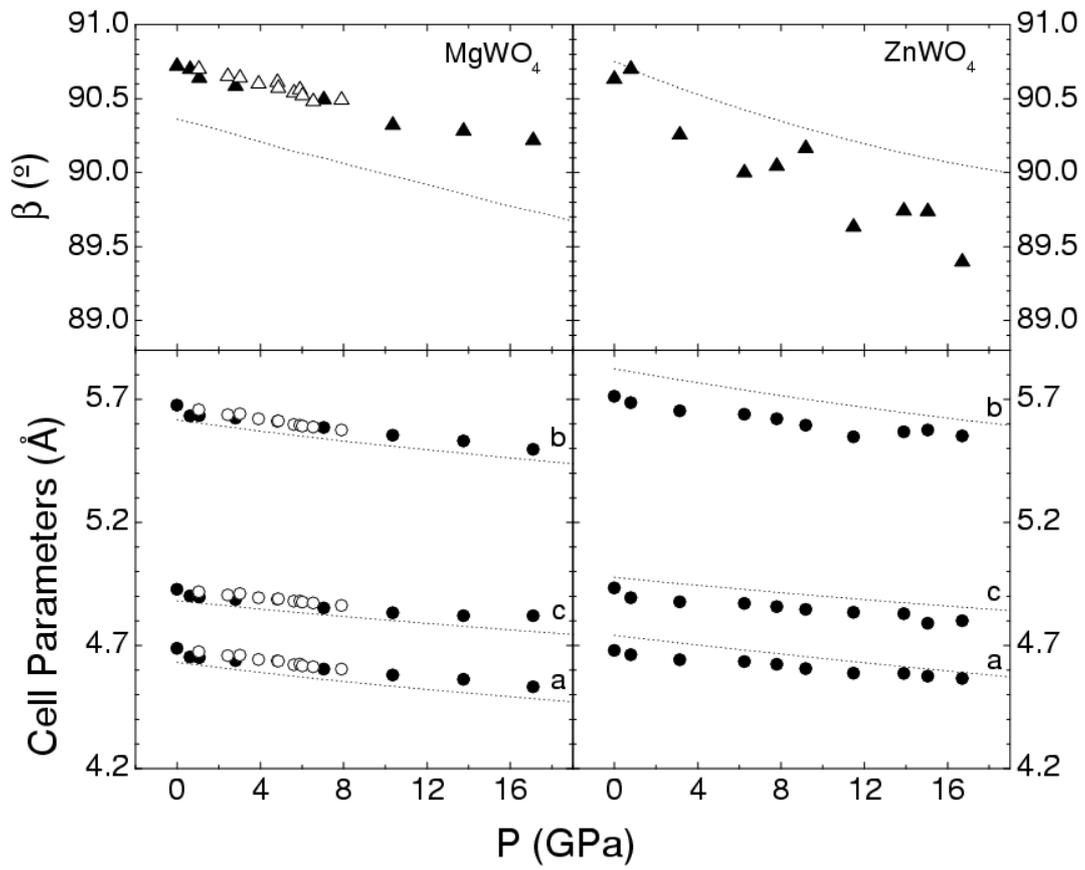

**Figure 7**

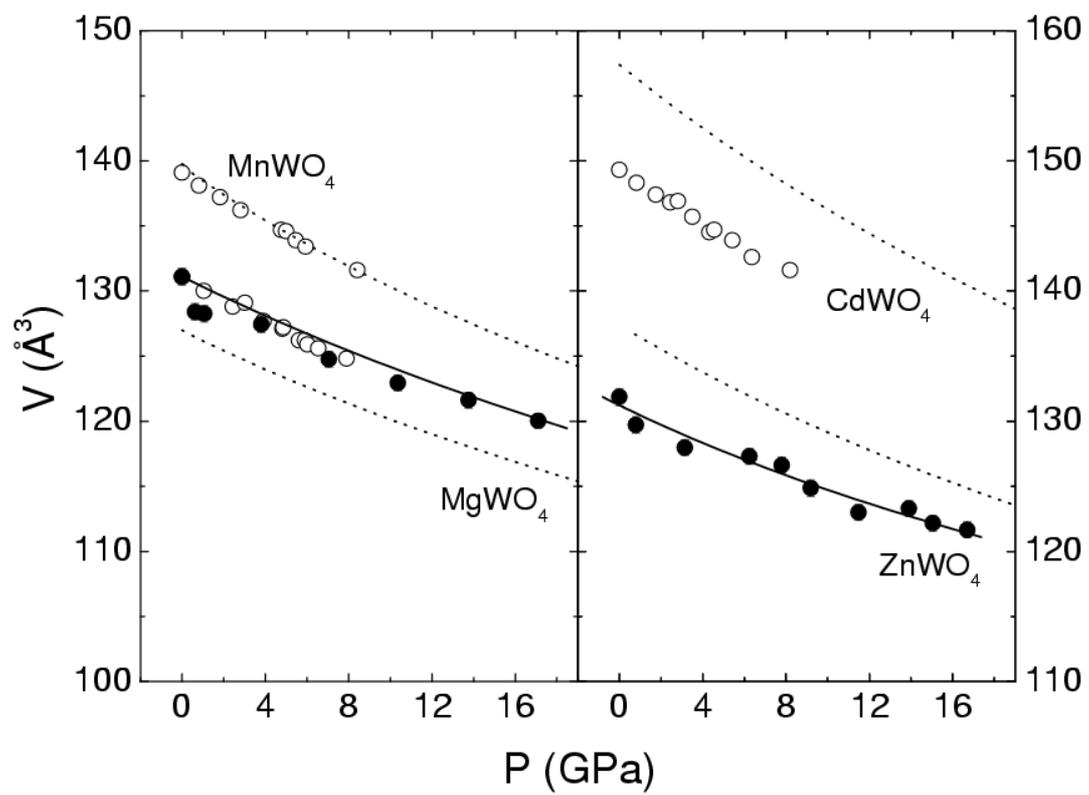

**Figure 8**

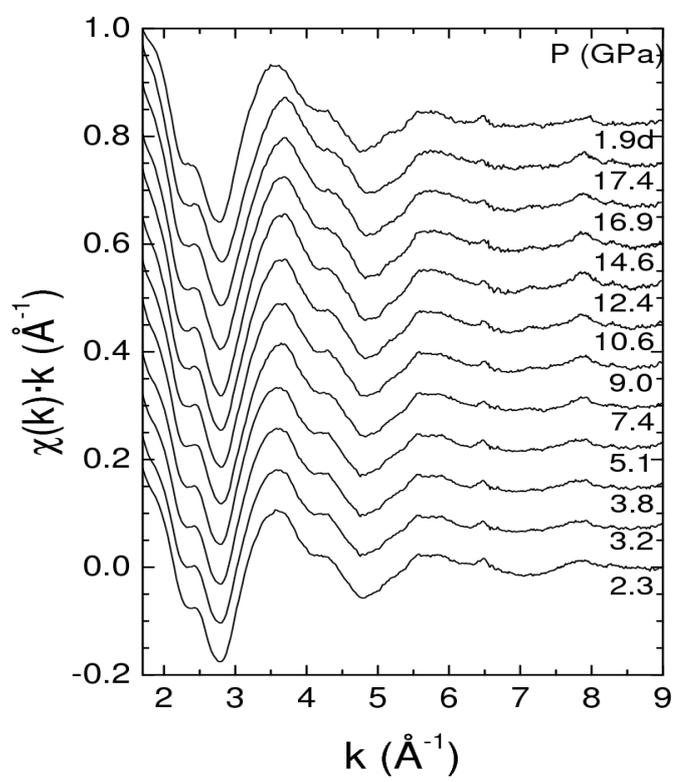

**Figure 9**

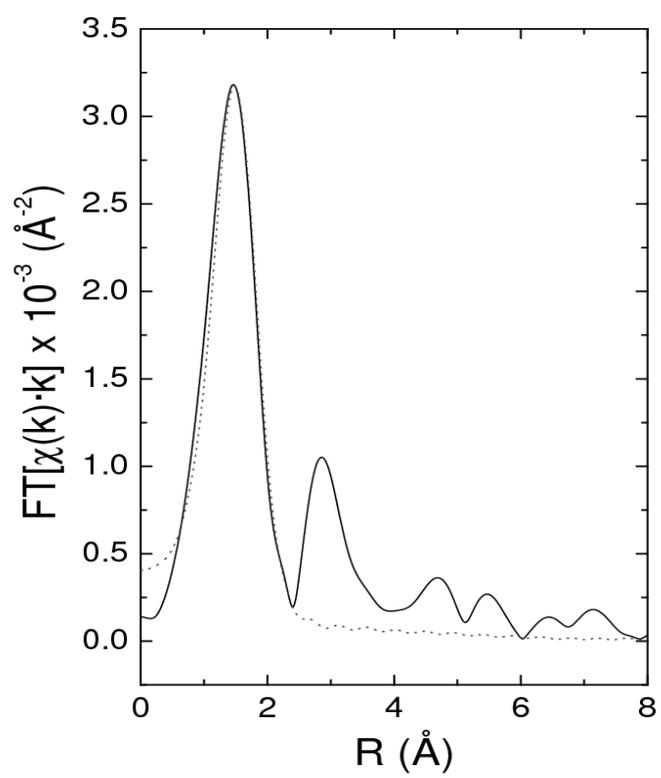

**Figure 10**

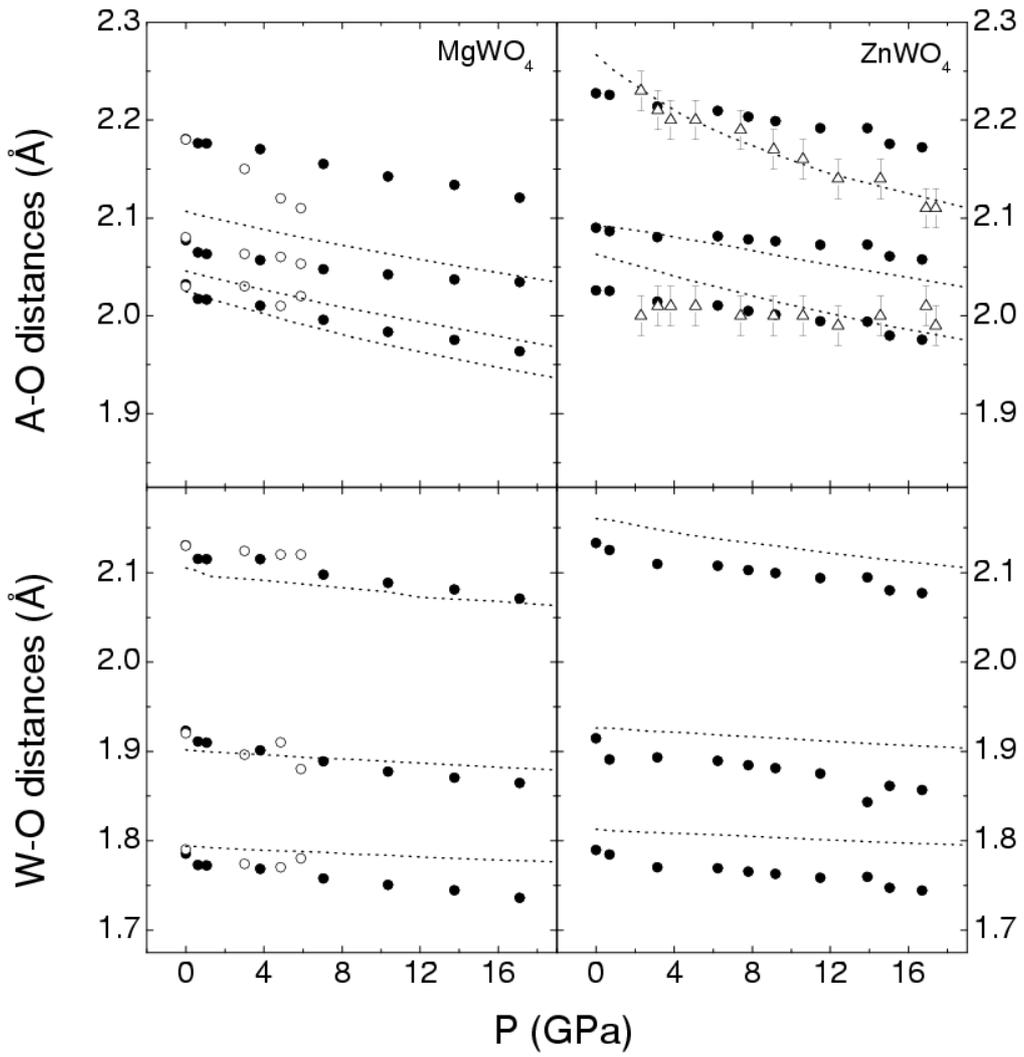

**Figure 11**

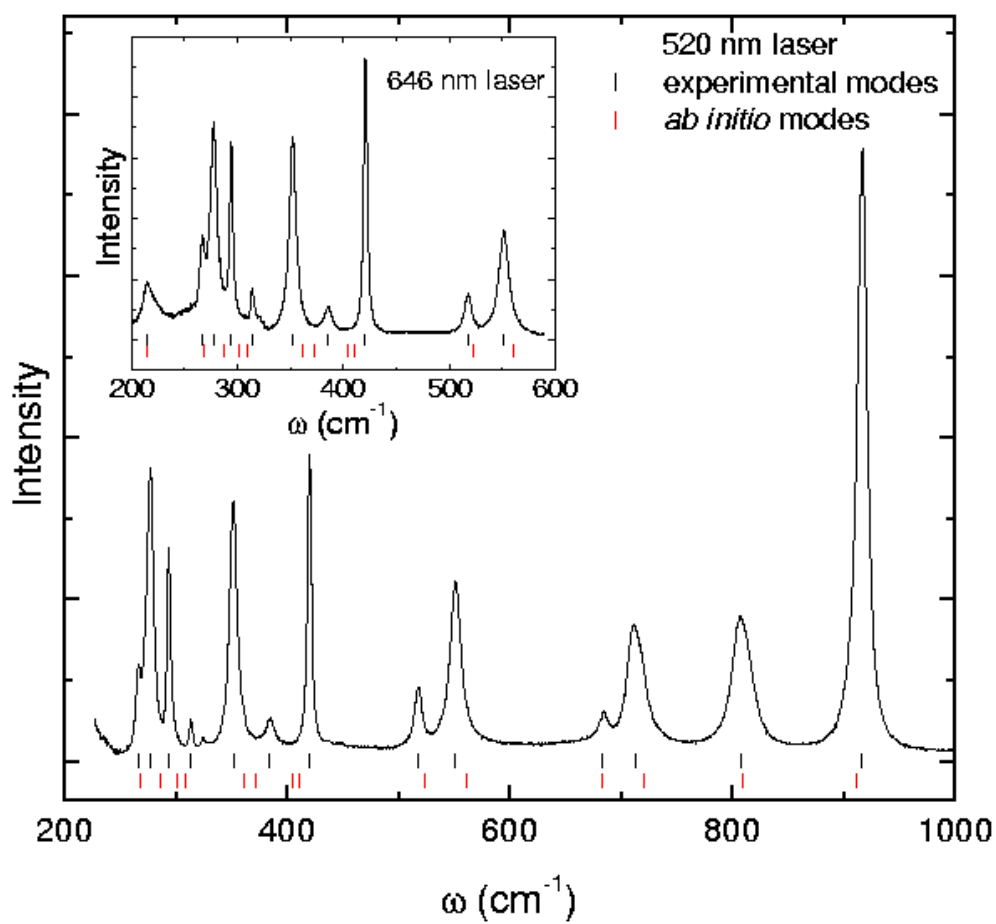

**Figure 12**

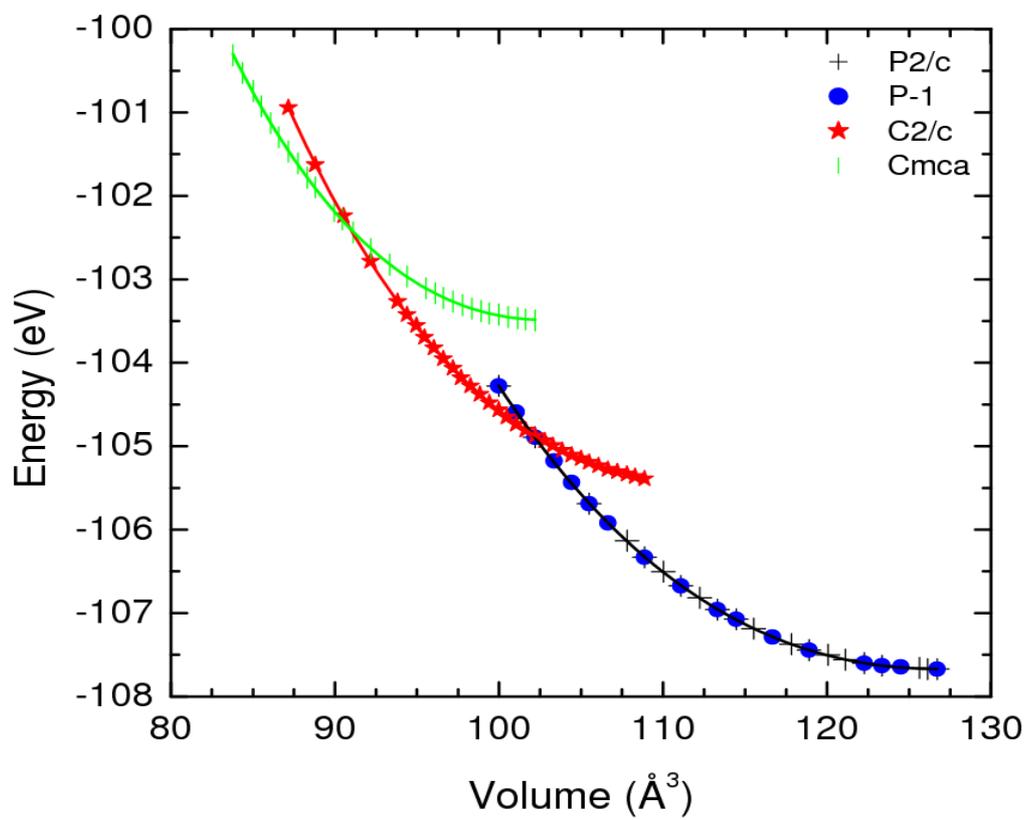